\documentclass[aps,twoside]{revtex4}
\usepackage{epsfig}
\usepackage{graphicx}
\usepackage{amsmath,amssymb,color}
\usepackage[english]{babel}

\parskip=\medskipamount

\newcommand{\eq}[1]{(\ref{#1})}
\newcommand{\fig}[1]{Fig.~\ref{#1}}

\newcommand{\be}{\begin{equation}}
\newcommand{\ee}{\end{equation}}

\newcommand\disp{\displaystyle}

\newcommand{\ve}{\mathbf}

\def\runninghead#1#2{\pagestyle{myheadings}
\markboth{{\protect\it{\quad #1}}\hfill} {\hfill{\protect\it{#2\quad}}}}

\runninghead{M Tamm, N. Pospelov, and S. Nechaev}{Growth rate of 3D Tetris}

\begin{document}

\title{Growth rate of 3D heaps of pieces}

\author{M.V. Tamm$^{1,2}$, N. Pospelov$^{1}$, and S. Nechaev$^{3,4}$}

\affiliation{$^1$ Faculty of Physics, Lomonosov Moscow State University, 119992 Moscow, Russia \\ $^2$ Department of Applied Mathematics, MIEM, National Research University Higher School of Economics, 123458, Moscow, Russia \\ $^3$ Interdisciplinary Scientific Center Poncelet (CNRS UMI 2615), 119002 Moscow, Russia \\ $^4$ P.N. Lebedev Physical Institute RAS, 119991 Moscow, Russia}

\begin{abstract}

We consider configurational statistics of three-dimensional heaps of $N$ pieces ($N\gg 1$) on a simple cubic lattice in a large 3D bounding box of base $n \times n$, and calculate the growth rate, $\Lambda(n)$, of the corresponding partition function, $Z_N\sim N^{\theta}[\Lambda(n)]^N$, at $n\gg 1$. Our computations rely on a theorem of G.X. Viennot \cite{viennot-rev}, which connects the  generating function of a $(D+1)$-dimensional heap of pieces to the generating function of projection of these pieces onto a $D$-dimensional subspace. The growth rate of a heap of cubic blocks, which cannot touch each other by vertical faces, is thus related to the position of zeros of the partition function describing 2D lattice gas of hard squares. We study the corresponding partition function exactly at low densities on finite $n\times n$ lattice of arbitrary $n$, and  extrapolate its behavior to the jamming transition density. This allows us to estimate the limiting growth rate, $\Lambda =\lim_{n\to\infty}\Lambda(n)\approx 9.5$. The same method works for any underlying 2D lattice and for various shapes of pieces: flat vertical squares, mapped to an ensemble of repulsive dimers, dominoes mapped to an ensemble of rectangles with hard-core repulsion, etc.

\end{abstract}

\maketitle

\section{Introduction}

We propose an estimation of the growth rate of a family of three-dimensional heaps of pieces (HP) which is reminiscent of the famous ``Tetris'' game on a simple cubic lattice, in which pieces of various shapes are dropped down along vertical direction until they hit already deposited elements. Formally, a heap of pieces is a collection of elementary blocks which are piled together. The main goal of the HP problem consists in computing the asymptotic behavior of the partition function, $Z_n(N) \sim N^{\theta}[\Lambda(n)]^N$, of all allowed distinct configurations of $N$-particle heaps ($N\to \infty$) in a three-dimensional bounding box with the base $n\times n$ on a lattice with a given symmetry. The critical exponent, $\theta$, is universal and depends only on the space dimensionality, while $\Lambda(n)$ is sensitive to the lattice geometry and interaction between pieces.

\begin{figure}[ht]
\includegraphics[width=12cm]{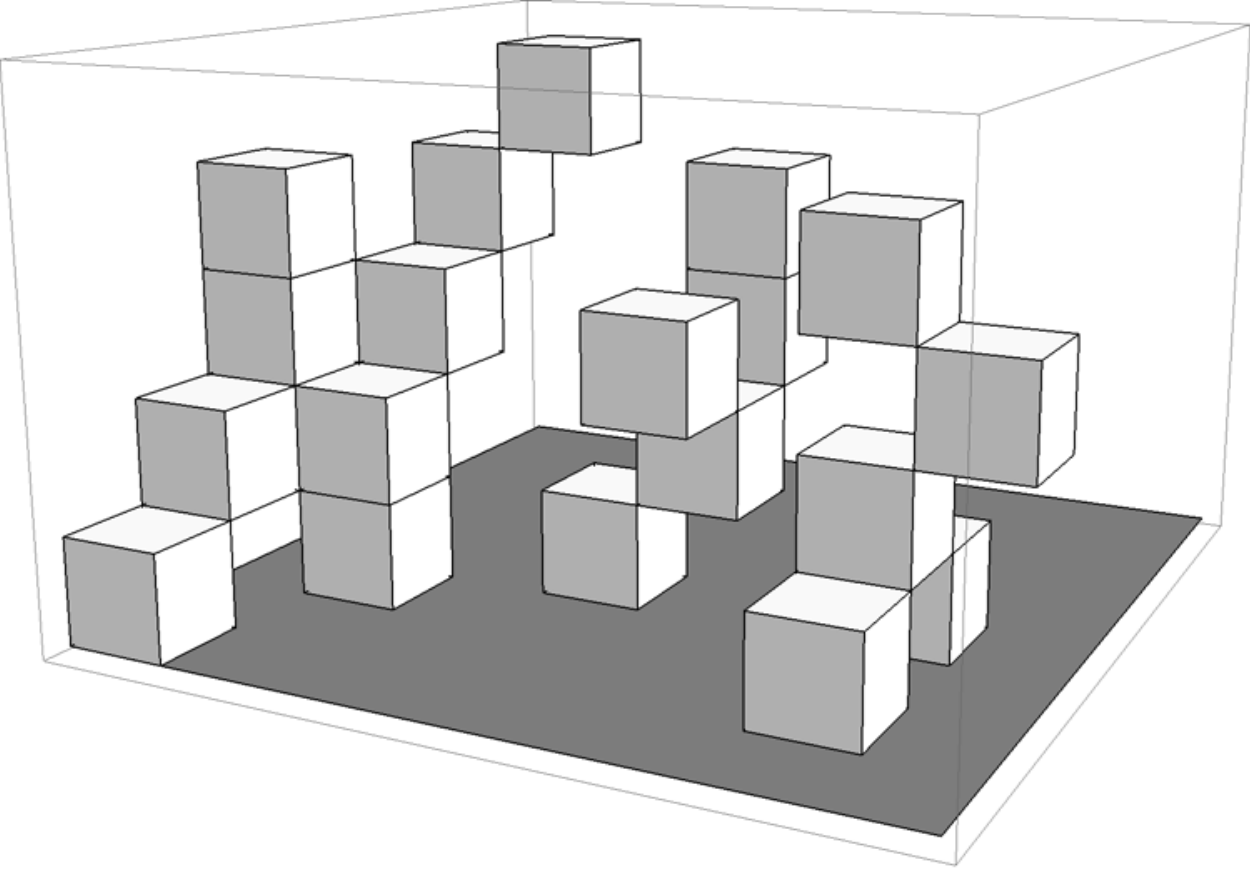}
\caption{A typical configuration of a three-dimensional heap of cubic blocks on a simple cubic lattice.}
\label{fig:01}
\end{figure}

We mostly discuss the heaps of cubic pieces which cannot touch each other by their side faces (see \fig{fig:01} for a typical configuration of such a heap). Although \fig{fig:01} is  self-explanatory, it seems instructive to define the precise rules of the heap construction. Take an empty 3D bounding box with the square base (``the floor'') consisting of $n\times n$ elementary squares of size $1\times 1$. Add cubic blocks of size $1\times 1\times 1$ according to the following rules. First, distribute any number of blocks on the ``floor'' (with the vertical coordinate 0) in a way that blocks have no common faces (common edges are allowed). The resulting set of blocks (the first layer) we call the ``roots'' of the heap. Now, add new blocks to the second layer (i.e. layer with the vertical coordinate 1). The blocks of this layer must fulfil two requirements: (i) they cannot have common faces with each other, and (ii) each of them must be a nearest neighbor to one of the roots, i.e. be located either exactly on top of a root block, or have one common edge with it. Further layers are added recursively with a preceding layer regarded as a root. Thus, any block in the heap can be reached from at least one root along a \emph{directed} path running through the collection of touching blocks (the path is called directed if it connects layers in a sequentially increasing order).

Note that the system described above might be thought of as a result of some deposition process with pieces falling down from $z=+\infty$ until they reach the lowest possible position respecting the constraint that no pieces have common vertical faces. Indeed, one and only one heap can be constructed in such a way. However, any possible deposition rules impose a probability measure over heaps: in a random deposition process some heaps are obtained more often than the others. In that sense the problem discussed here is not a deposition problem: here we consider all different heaps as equiprobable, and are interested in their total number.

The same combinatorial problem can be posed in any dimension $D$ and for any graph as a floor. The concept of a heap of pieces has appeared apparently for the first time in 1969 in the work of P. Cartier and D. Foata \cite{cartier} where they considered monoids generated by some alphabet with the relations $ab = ba$. Various related models, as well as new combinatorial results and the relation with the statistical physics is reviewd in \cite{viennot-rev}. The 2D heaps of pieces on square and triangular lattices have been exhaustively studied in the literature and played a role of a testing ground for various approaches -- from purely combinatorial \cite{viennot-rev, betrema, bousquet3}, to the ones based on the diagonalization of a special transfer matrix and Bethe ansatz computations \cite{hakim1,dhar1,dhar2,dhar}.

Not much is known about the partition functions of three-dimensional heaps of pieces. To the best of our knowledge, the only one known solution of the 3D HP problem was published in 1983 by D. Dhar in \cite{dhar} by an exact mapping of the HP on a body-centered-cubic lattice onto the two-dimensional hard hexagon model. The solution of the latter system has been obtained by R. Baxter using the machinery of the Bethe ansatz \cite{baxter}.

The HP problem met essential attention of both physicists and mathematicians. Besides the physics of growing clusters, several other problems in pure mathematics and mathematical physics are connected to the HP model. For example, various aspects \cite{hakim1, viennot1, bousquet1, bousquet2} of the enumerative combinatorics of partitions are related to growth of 2D heaps of pieces. In \cite{vershik} the statistics of growing heaps has been linked to the statistics of two-dimensional growing braids; in \cite{anim-math} the general asymptotic theory of directed two-dimensional lattice paths in half-planes and quarter-planes has been reviewed; and in \cite{haug} the algebraic ansatz for the steady-state of the totally asymmetric simple exclusion process (TASEP) on an open line has been interpreted in terms of combinatorics of 2D heaps of pieces. The discussion of algebraic approach to the ASEP model can be found in a seminal paper \cite{evans}.

In our work we compute the growth rate, $\Lambda(n)$, of the ensemble of 3D heaps of cubic blocks on a simple cubic lattice in a bounding box of base $n\times n$. The growth rate $\Lambda(n)$ defines the asymptotic behavior of the partition function, $Z_n(N)$:
\be
\lim_{N\to\infty} \frac{\ln Z_n(N)}{N} =\ln \Lambda(n)
\label{e:01}
\ee
Our approach relies on a beautiful relation of G. Viennot \cite{viennot-rev}, that does not seem to be very widely known, and which relates the partition function of a heaps of pieces to the partition function of disposition of the projection of these pieces onto the floor of the heap. We combine that relation with the semi-phenomenological analysis of zeroes of the 2D (``floor'') partition function in order to obtain an estimate for the limiting growth rate, $\Lambda = \lim_{n\to\infty} \Lambda(n)$.

\section{HP partition function on a simple cubic lattice}

\begin{figure}[ht]
\includegraphics[width=6cm]{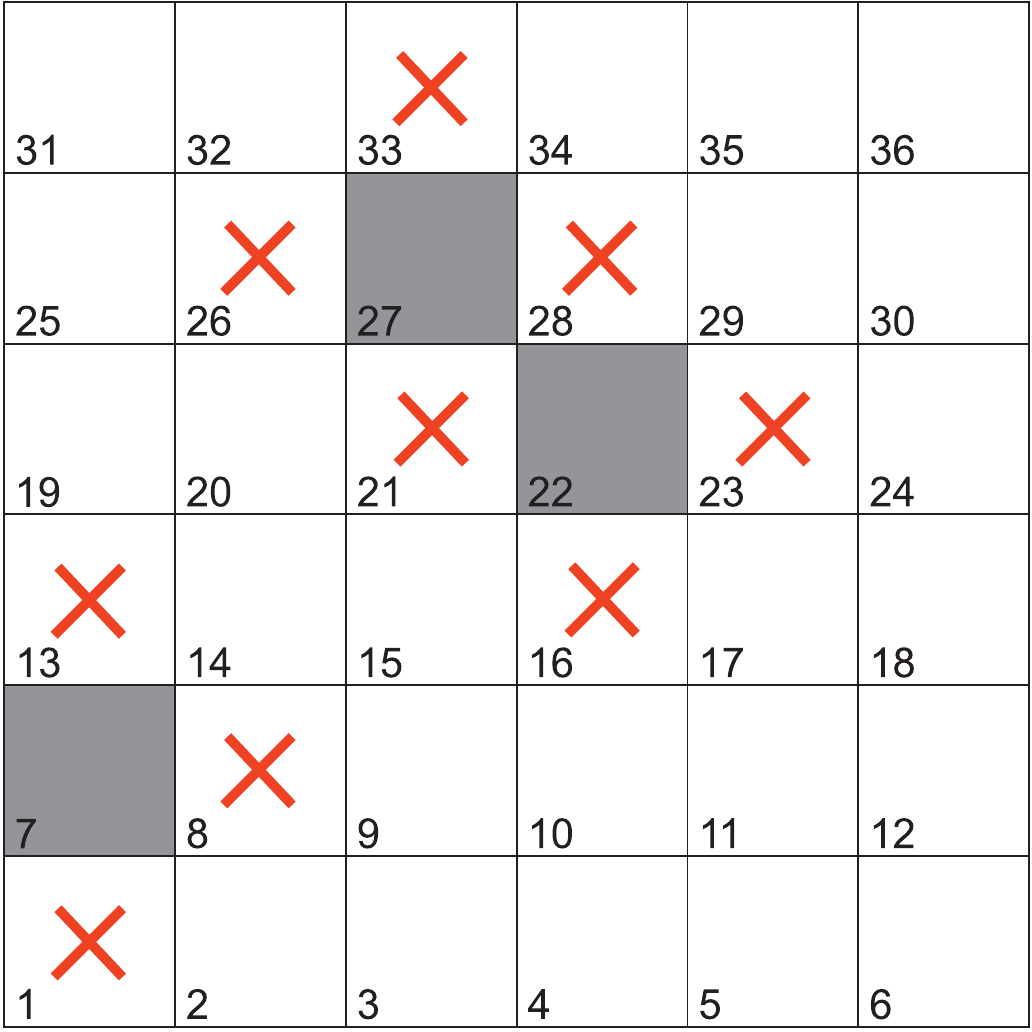}
\caption{A configuration of 3 squares with hard-core interactions on a $6\times 6$ square lattice. Particles are grey squares, red crosses mark the positions forbidden for the particles.}
\label{fig:02}
\end{figure}

We start with reminder of G. Viennot theorem \cite{viennot-rev}, which is a keystone of our consideration. Let $Z_n(N)$ be the partition function (total number of configurations) of $N$-particle heap of nearest-neighbor interacting cubic blocks (see \fig{fig:01}) inside a bounding box with $n\times n$ base, and $G_n(s)$ is the corresponding generating function (grand canonical partition function):
\be
G_n(s) = \sum_{N=1}^{\infty} Z_n(N) s^N \equiv \sum_{\text{allowed configurations}} s^{\# \text{ of blocks in configuration}}
\label{e:02}
\ee
Now, consider the set of all possible roots of a heap. That is to say, we take a two-dimensional $n\times n$ square lattice, on which we distribute $1\times 1$ squares in a way that they cannot have common sides (see \fig{fig:02} for a typical configuration with $n=6$). Let $A_{k,n}$ be the total number of configurations of $k$ particles on the $n\times n$-square, and $W_n(s)$ -- the corresponding generating polynomial (grand canonical partition function of a square lattice gas with the hard core repulsion):
\be
W_n(s) = 1 + \sum_{k=1}^{M_{\max}} A_{k,n} s^k
\label{e:03}
\ee
by $M_{\max} (n)$ we denoted the maximal number of particles which can be arranged on a $n\times n$ lattice,
\be
M_{\max} =\left[\frac{n^2+1}{2}\right]
\label{e:04}
\ee

The theorem of G. Viennot \cite{viennot-rev} states that there is a following relation between $G_n(s)$ and $W_n(s)$:
\be
G_n(s) = \frac{1}{W_n(-s)}\equiv \frac{1}{\disp 1+\sum_{k=1}^{M_{\max}} (-1)^k\, A_k\, s^k}.
\label{e:05}
\ee
For completeness, let us give here a sketch of the proof. Consider the product $G_n(s) W_n(t)$, which enumerates configurations in the direct product of: (a) the set $G_n(s)$ of \emph{all possible heaps}, whose pieces are enumerated by $s$, and (b) the set $W_n(t)$ of \emph{all possible single layers}, whose pieces are enumerated by $t$. In what follow we call a representative of the first set ``a heap of $s$-pieces'', and of the second set -- ``a plate of $t$-pieces''. Now, consider an element of $G_n(s) W_n(t)$, i.e., a pair of a heap and a plate, and think of them in a following way. Put the plate at the bottom floor, so that all $t$-pieces have vertical coordinate 0, and then put the heap on top of it. That is to say, first put it so that it feels planes with vertical coordinates 1, 2, 3, ..., and then allow pieces to fall down if underlying layer does not prevent them from doing so. Now, we can write down the generating function of all the resulting structures in a following way
\be
G_n(s) W_n(t) = \sum_{\alpha} t^{n_\alpha} \sum_{\beta}s^{n_\beta} F_{\alpha,\beta}(s),
\ee
where $\alpha$ and $\beta$ enumerate all possible configurations of $t$-pieces and $s$-pieces in the lowest layer (combination of $t$-pieces being the original plate, and combination of $s$-pieces -- whatever fell down from the upper layer as a result of a procedure described above), $n_{\alpha,\beta}$ are the respective numbers of pieces in the lowest layer, and $F_{\alpha,\beta}(s)$ is the generating function of all heaps that can be placed on top of that fixed lowest layer configuration. Note now that  $F_{\alpha,\beta}(s)$ is in fact a function of only the \emph{total} configuration of the lowest layer, $\alpha \cup \beta$, and not of the way, how the pieces are separated into $s$-type and $t$-type. Therefore,
\be
G_n(s) W_n(t) = \sum_{\alpha\cup\beta} F_{\alpha \cup \beta}(s) \sum_{\alpha} t^{n_\alpha} s^{n_\beta} =  \sum_{\alpha\cup\beta} F_{\alpha \cup \beta}(s) (s+t)^{n_{\alpha \cup \beta}} ,
\ee
where the first sum is over all possible configurations of the lowest layer, and the second -- over all possible separations of its pieces into $s$- and $t$-types. The last equation allows for the fact that each piece can be assigned to either $s$- or $t$-type indpeendently of others. Thus, at $t = -s$ only the term with $n_{\alpha \cup \beta}=0$ (i.e. which corresponds to empty layer and empty heap) survives in the sum, and therefore
\be
G_n(s) W_n(-s) = 1,
\ee
proving the theorem of G. Viennot \cite{viennot-rev}.

Now, $Z_n(N)$ can be obtained from $G_n(s)$ in a usual way
\be
Z_n(N) = \frac{1}{2\pi i} \oint \frac{G_n(s)}{s^{n+1}}ds
\label{e:06}
\ee
and, therefore, the growth rate of $Z_n$ is simply
\be
\Lambda(n) = -s_*^{-1},
\label{lambda}
\ee
where $s_*$ is zero of the $W_n(s)$ polynomial with the smallest absolute value.

For illustrative purposes in Appendix \ref{app1} we calculate $G_n(s)$ and $W_n(s)$ explicitly for $n=2,3$ and check directly that they do satisfy \eq{e:05}.

By \eq{e:05} the combinatorics of 3D heaps of pieces is reformulated as a problem of calculating the canonical partition function of the lattice ``hard-square gas'', which in turn can be mapped onto the law-temperature limit of the canonical 2D Ising model on a finite square lattice in a constant magnetic field. Since exact solution of the 2D Ising model in the magnetic field is unknown, one has to rely on heuristic arguments concerning the behavior of zeroes of $W_n(-s)$, which allow us to conjecture the growth rate of the 3D heap of cubic blocks in a large bounding box (at $n\gg 1$).

We have calculated, by means of direct enumeration, all coefficients $A_{k,n}$ in \eq{e:03} for $n = 2...7$, and have got the following exact explicit expressions for the polynomials $W_2(s),...,W_7(s)$:
\be
\begin{array}{ll}
W_2(s) = 1 & \hspace{-0.2cm} + 4 s + 2 s^2; \medskip \\
W_3(s) = 1 & \hspace{-0.2cm} + 9 s + 24 s^2 + 22 s^3 + 6 s^4 + s^5;  \medskip \\
W_4(s) = 1 & \hspace{-0.2cm} + 16 s + 96 s^2 + 276 s^3 + 405 s^4 + 304 s^5 + 114 s^6 + 20 s^7 +
 2 s^8; \medskip \\
W_5(s) = 1 & \hspace{-0.2cm} + 25 s + 260 s^2 + 1474 s^3 + 5024 s^4 + 10741 s^5 + 14650 s^6 +
 12798 s^7 + 7157 s^8 \\ & \hspace{-0.2cm} + 2578 s^9 + 618 s^{10} + 106 s^{11} +
 14 s^{12} + s^{13}; \medskip \\
W_6(s) = 1 & \hspace{-0.2cm} + 36 s + 570 s^2 + 5248 s^3 + 31320 s^4 + 127960 s^5 + 368868 s^6 +
 763144 s^7 + 1143638 s^8 \\ & \hspace{-0.2cm} + 1247116 s^9 + 991750 s^{10} + 576052 s^{11} +
 245030 s^{12} + 76716 s^{13} + 17834 s^{14} + 3120 s^{15} \\ & \hspace{-0.2cm} + 416 s^{16} +
 40 s^{17} + 2 s^{18}; \medskip \\
W_7(s) = 1 & \hspace{-0.2cm} + 49 s + 1092 s^2 + 14690 s^3 + 133544 s^4 + 870589 s^5 +
 4216498 s^6 + 15516804 s^7 + 44031035 s^8 \\ & \hspace{-0.2cm} + 97284860 s^9 +
 168434824 s^{10} + 229465420 s^{11} + 246654832 s^{12} + 209621800 s^{13} +
 141138510 s^{14} \\ & \hspace{-0.2cm} + 75497964 s^{15} + 32229088 s^{16} + 11059218 s^{17} +
 3084524 s^{18} + 710124 s^{19} + 137368 s^{20} + 22611 s^{21} \\ & \hspace{-0.2cm} + 3134 s^{22} +
 344 s^{23} + 26 s^{24} + s^{25};
\end{array}
\label{e:07}
\ee

\begin{figure}[ht]
\includegraphics[width=16cm]{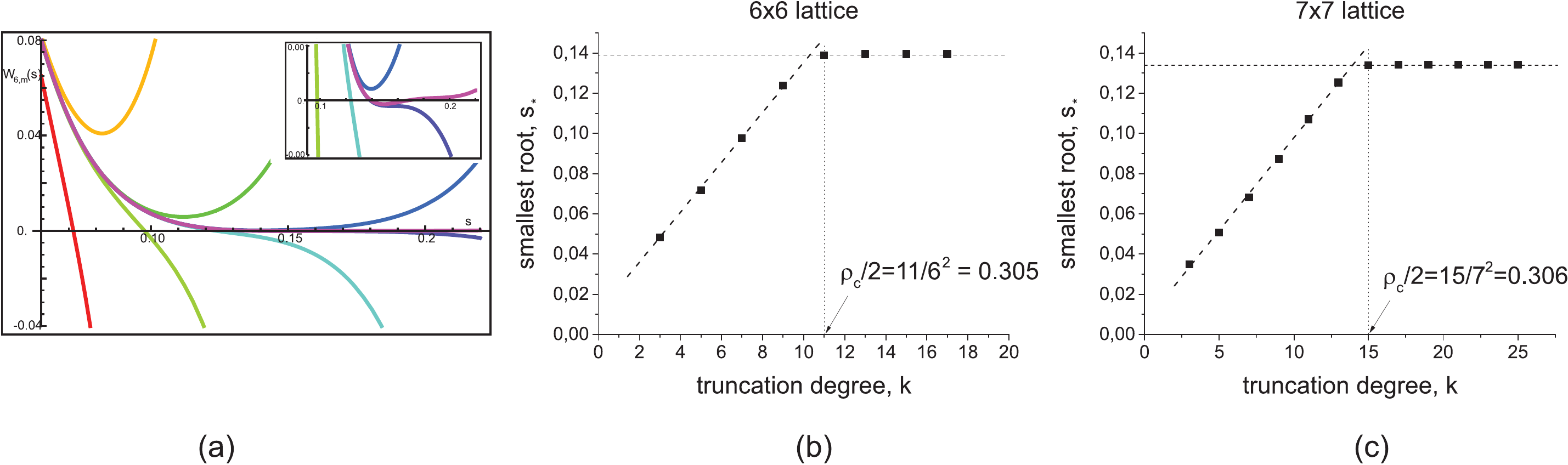}
\caption{Behavior of smallest real roots of truncated polynomials $\overline{W}_{n,m}(-s)$: (a) truncated polynomials $\overline{W}_{6,m}(-s)$ as a function of $s$ for $m=5...12$ (colored from red to purple with increasing $m$), inset shows the vicinity of $\overline{W}=0$; (b,c) positions of smallest real roots as a function of $m$ for $n=6$ (b) and $7$ (c).}
\label{fig:03}
\end{figure}

As it is stated by \eq{lambda}, the growth rate of the 3D heap $\Lambda(n)$ is determined by the position of the root $s_*(n)$ of the polynomial $W_n(-s)$ with the smallest absolute value. In order to get some insight into the behavior of the smallest root, we have studied the roots of truncated polynomials $\overline{W}_{n,m}(-s)$, i.e., the polynomials with only $m$ first terms preserved. The smallest root of $\overline{W}_{n,m}(-s)$ is behaving in a very peculiar way as a function of $m$ -- see \fig{fig:03}. In \fig{fig:03}a  we draw the sequence of truncated polynomials $\overline{W}_{n,m}(-s)$ for $n=6,7$ and different $m$. One sees that for small even $m$ the polynomials have no real roots, while for small odd $m$ the smallest root exhibits nearly linear growth as it is shown in \fig{fig:03}. Then, at some critical $m_c$ a real root of the polynomial shows up for even $m$ as well, and starting from $m_c$ the root stabilizes, so that further increase of $m$ does not change the root.

Based on that observation, we suggest the following heuristic procedure of estimating the value of the smallest root $s_*(n)$ on the basis of the behavior $m_c(n)$:

\noindent 1. From \fig{fig:03}b,c we expect that the fraction of occupied cites
\be
\rho_c = \frac{m_c}{M_{\max}}= \frac{m_c}{\left[\frac{n^2+1}{2}\right]}
\label{rho}
\ee
is approximately constant as a function of $n$. It seems reasonable to identify $\rho_c$ with the density of ordering transition in the 2-dimensional square lattice gas with the hard-core repulsion \cite{burley, Fisher65}. Analyzing works \cite{burley,Fisher65,Runnels,Baxter2,blote,fixman,mean}, one can conclude that the phase transition in the 2D hard square gas lies in the interval $0.61<\rho_{c}<0.75$, and $\rho_{c}\approx 0.728$ is the most probable transition value.

\noindent 2. Assume some functional dependence $s_0(m)$ for $m<m_c$, where $m$ is the truncation degree of the polynomial. It can be seen from \fig{fig:03}b,c that the behavior is almost perfectly linear for $n=6,7$ (it is also true for all smaller $n$), so we suggest the linear dependence
\be
s_0(m|n)= A(n) + B(n)\; m \qquad (\mbox{for odd $m \le m_{c}$})
\label{07b}
\ee
Now, given $A(n)$ and $B(n)$ one can estimate the limiting value of the root $s_*(n)$ by substituting $m_c = \rho_c M_{\max}$ into \eq{07b}. In the next section we calculate explicitly several first terms of the polynomials $W_n(s)$ for arbitrary $n$, which allows us to get estimates of $A(n)$ and $B(n)$ and, thus, of $s_*(n)$.

\section{Evaluation of the coefficients $A_{k,n}$ for hard-square lattice gas}

In order to evaluate the first terms in $W_n(s)$ one proceeds as follows. Enumerate the plaquettes of the $n\times n$ square lattice sequentially row-by-row as shown in \fig{fig:02}, and introduce a $n^2 \times n^2$ ``incidence'' matrix $T$ with matrix elements $t_{ij}$, encoding the allowed configurations of filled plaquettes. Namely,
\be
t_{ij} = \begin{cases} 0 & \mbox{if $i=j$ or $i$ and $j$ are nearest neighbours} \medskip \\ 1 & \mbox{otherwise} \end{cases}
\label{e:T}
\ee
For example, for $n=4$ the matrix takes form
\be
T=\left(
\begin{array}{cccccccccccccccc}
  0 & 0 & 1 & 1& 0 & 1 & 1 & 1 & 1 & 1 & 1 & 1 & 1 & 1 & 1 & 1  \\
  0 & 0 & 0 & 1 & 1 & 0 & 1 & 1 & 1 & 1 & 1 & 1 & 1 & 1 & 1 & 1  \\
  1 & 0 & 0 & 0 & 1 &1 &  0 & 1 & 1 & 1 & 1 & 1 & 1 & 1 & 1 & 1  \\
  1 & 1 & 0 & 0 & 1 &1 &  1 & 0 & 1 & 1 & 1 & 1 & 1 & 1 & 1 & 1  \\
   0 & 1& 1 & 1 & 0 & 0 & 1 & 1 & 0 & 1 & 1 & 1 & 1 & 1 & 1 & 1  \\
  1 & 0 & 1 & 1& 0 & 0 & 0 &1 &  1 & 0 & 1 & 1 & 1 & 1 & 1 & 1  \\
  1 & 1 & 0 & 1& 1 & 0 & 0 & 0 &1 &  1 & 0 & 1 & 1 & 1 & 1 & 1  \\
  1 & 1 & 1 & 0& 1 & 1 & 0 & 0 & 1 & 1 & 1 & 0 & 1 & 1 & 1 & 1  \\
  1 & 1 & 1 & 1 & 0& 1 & 1 & 1 & 0 & 0 & 1 & 1 & 0 & 1 & 1 & 1  \\
  1 & 1 & 1 & 1 & 1 & 0& 1 & 1 & 0 & 0 & 0 &1 &  1 & 0 & 1 & 1  \\
  1 & 1 & 1 & 1 & 1 & 1 & 0& 1 & 1 & 0 & 0 & 0 & 1 & 1 & 0 & 1  \\
  1 & 1 & 1 & 1 & 1 & 1 & 1 & 0& 1 & 1 & 0 & 0 & 1 & 1 & 1 & 0  \\
  1 & 1 & 1 & 1 & 1 & 1 & 1 & 1 & 0& 1 & 1 & 1 & 0 & 0 & 1 & 1  \\
  1 & 1 & 1 & 1 & 1 & 1 & 1 & 1 & 1 & 0& 1 & 1 & 0 & 0 & 0 & 1 \\
  1 & 1 & 1 & 1 & 1 & 1 & 1 & 1 & 1 & 1 & 0& 1 & 1 & 0 & 0 & 0 \\
  1 & 1 & 1 & 1 & 1 & 1 & 1 & 1 & 1 & 1 & 1 & 0& 1 & 1 & 0 & 0
\end{array}
\right),
\label{e:11}
\ee
where $0$'s on the first subdiagonal correspond to horizontally adjacent squares (there are gaps when enumeration jumps from one raw to another), and $0$'s on the fourth subdiagonal -- to the vertically adjacent ones. Then for any $k \geq 1$ the coefficient $A_{n,k}$ in the expansion of $W_n(s)$, which is just the number of different ways to arrange $k$ squares on the $n\times n$ lattice respecting the pairwise conditions encoded in the matrix $T$, can be formally written as follows
\be
A_{n,k} =\frac{1}{k!} \sum_{i_1=1}^{n^2} \sum_{i_2=1}^{n^2}\dots \sum_{i_k=1}^{n^2}  \prod_{\alpha>\beta} t_{i_\alpha,i_\beta}
\label{A_general}
\ee
Indeed, here index $i_{\alpha}$ enumerates all possible positions of the $\alpha$'th particle, the product over all pairs of particles checks that none of them violates the nearest-neighboring restriction, and $k!$ allows for the permutations in particle enumeration.

For $k=1$ there is no product in the equation \eq{A_general}, so the result is simply
\be
A_{n,1} = \sum_{i=1}^{n^2} 1 = n^2
\label{A_1}
\ee

\begin{figure}[ht]
\includegraphics[width=14cm]{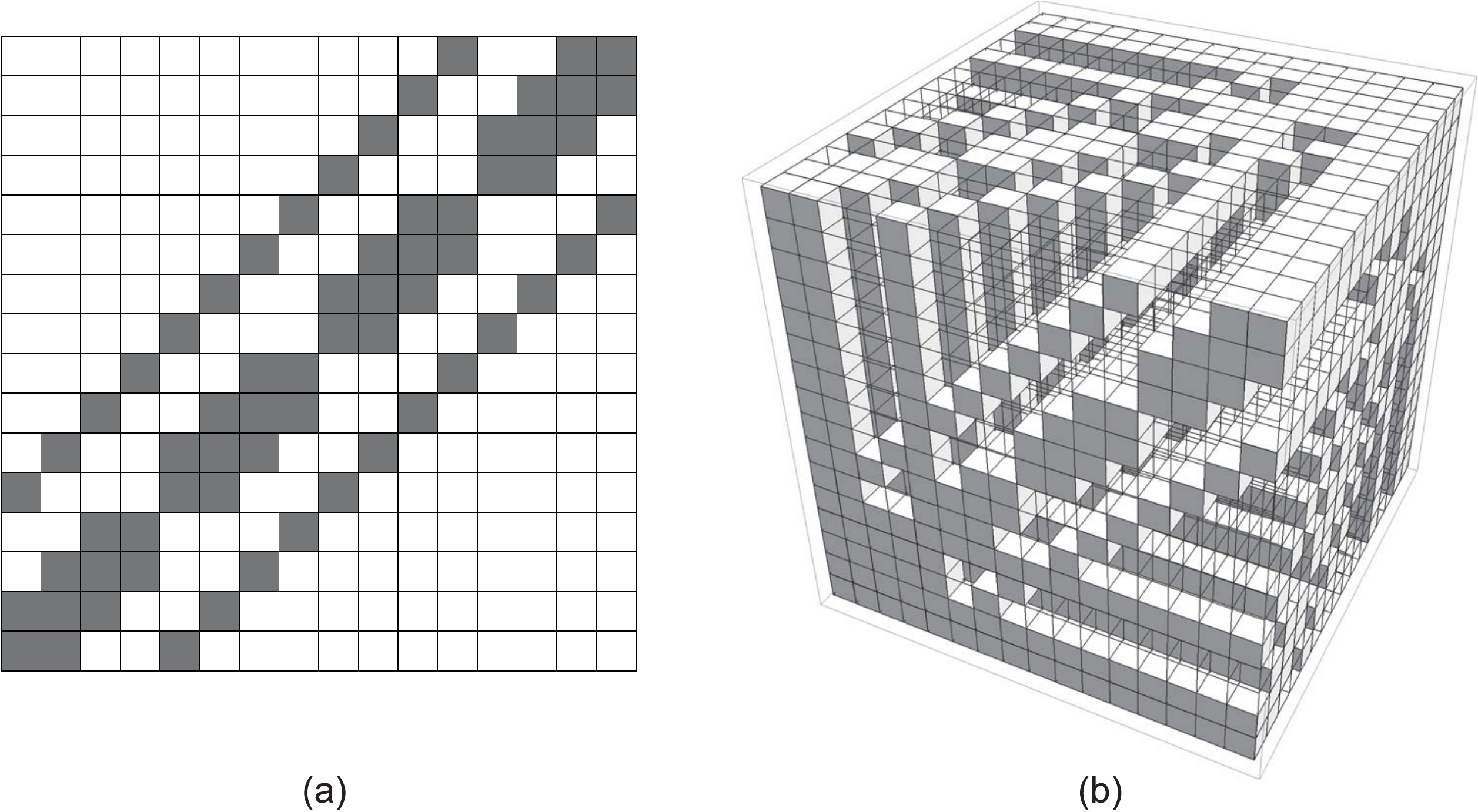}
\caption{Counting the number of available positions of 2 and 3 hard squares in the bounding box: (a) number of 2-particle configurations is equal to empty cells in the 2D array; (b) number of 3-particle configurations is equal to number of empty cells in the shown 3D pierced structure.}
\label{fig03}
\end{figure}

For $k=2$ the value $A_{n,2}$ in \eq{A_general} equals to half of total number of $1$'s in the matrix $T$, which is equal to the number of elementary white squares in the table shown in fig \fig{fig03}a:
\be
A_{n,2} = \frac{1}{2} \sum_{i=1}^{n^2} \sum_{j=1}^{n^2} t_{ij} = \frac{n^4 - n^2 - 4(n^2-n)}{2} = \frac{n^4 -5n^2+4n}{2}
\label{A_2}
\ee
For $k=3$ the equation \eq{A_general} reads
\be
A_{n,3} = \frac{1}{6} \sum_{i=1}^{n^2} \sum_{j=1}^{n^2} \sum_{k=1}^{n^2} t_{ij}t_{ik}t_{jk}
\label{A_3}
\ee
and corresponds to the enumeration of white cubes in the three-dimensional structure shown in \fig{fig03}b (for the  $n=4$ case). According to inclusion-exclusion principles, the sum \eq{A_3} can be written as
\be
A_{n,3} = \frac{1}{6} \left(a_0-a_1+a_2-a_3\right),
\label{A_3a}
\ee
It is easy to see that $a_0 = n^6$. One can explicitly calculate the number of black squares in \fig{fig03}a which is equal to $5n^2 - 4n$. Thus, $a_1 = 3n^2(5n^2-4n)$, where the coefficient 3 emerges from 3 unique pairs of axes: $(1,2), (2,3), (1,3)$. To obtain the terms $a_2$ and $a_3$, we turned to the polynomial fitting. The procedure is as follows:

\noindent (i) We explicitly calculate all $a_2$ and $a_3$ terms for 3-cubes of sizes $n=[3,20]$. That can be done by counting the number of elementary sites with values 2 or 3 in each 3-cube. It is useful to mention that the value of an elementary site $(i,j,k)$ in a cube is equal to the number of ``improperly placed pairs'' of sites on a square lattice $n \times n$.

\noindent (ii) We fit $a_i(n)$ by a polynomial with integer coefficients. Following this procedure, we obtained the expressions: $a_2 = 3(25n^2 - 36n + 8)$, $a_3 = 13n^2 - 12n$. Collecting all coefficients, we arrive at the final result for $A_{n,3}$:
\be
A_{n,3} = \frac{1}{6}(n^6 - 15n^4 + 12n^3 + 62n^2 - 96n + 24)
\label{e:15}
\ee

We succeeded in applying such way of reasoning further for calculating $A_{n,4}$ and $A_{5,n}$. Fortunately, with the help of \emph{Wolfram sequence solver} \cite{seq-sol} we were able to find \emph{On-line Encyclopedia of integer sequences} \cite{encyclopedia} and the book \cite{chess} where on page 370 the coefficients $A_{n,k}$ are obtained for $k=1,\dots,9$:
\be
\begin{array}{ll}
\disp A_{n,1} = n^2 & \medskip \\
\disp A_{n,2} = \frac{1}{2}(n^4-5n^2+4n) &  \medskip \\
\disp A_{n,3} = \frac{1}{3!}(n^6 -15n^4 + 12n^3 + 62n^2 - 96n + 24) & n \ge 2 \medskip \\
\disp A_{n,4} = \frac{1}{4!}(n^8 - 30n^6 + 24n^5 + 323n^4 - 504n^3 - 1110n^2 + 2760n - 1224) & n\ge 3 \medskip \\
\disp A_{n,5} = \frac{1}{5!} (n^{10} - 50n^8 + 40n^7 + 995n^6 - 1560n^5 - 8890n^4 + 21080n^3 + 24264n^2 - 97440n \\ \hspace{2cm} + 59520) & n \ge 4 \medskip \\
\disp  A_{n,6} = \frac{1}{6!}(n^{12} - 75n^{10} + 60n^9 + 2365n^8 - 3720n^7 - 38085n^6 + 89580n^5 + 292834n^4 - 984960n^3 \\ \hspace{2cm} - 552240n^2 + 4128960n - 3160800) & n \ge 5 \medskip \\
\disp A_{n,7} = \frac{1}{7!}(n^{14} - 105 n^{12} + 84 n^{11} + 4795 n^{10} - 7560 n^9 - 119595 n^8 + 280980 n^7+ 1660204 n^6 \\ \hspace{2cm} - 5360880 n^5 - 10985940 n^4 + 52150896 n^3 + 7858080 n^2 - 205168320 n + 187629120) & n \ge 6 \medskip \\
\disp A_{n,8} = \frac{1}{8!}(n^{16} - 140 n^{14} + 112 n^{13} + 8722 n^{12} - 13776 n^{11} - 309176 n^{10} + 726880 n^9 + 6592369 n^8 \\ \hspace{2cm} - 20974800 n^7 - 80804780 n^6 + 347456368 n^5 + 445614588 n^4 - 3115161504 n^3 \\ \hspace{2cm} + 533820336 n^2 + 11725156800 n - 12451057920) & n \ge 7 \medskip \\
\disp A_{n,9} = \frac{1}{9!}(n^{18} - 180 n^{16} + 144 n^{15} + 14658 n^{14} - 23184 n^{13} - 698040 n^{12} + 1643040 n^{11} \\ \hspace{2cm} + 20950545 n^{10} - 66240720 n^9 - 395401860 n^8 + 1635063696 n^7 + 4294617452 n^6 \\ \hspace{2cm} - 24611088096 n^5 - 17985223440 n^4 + 208162486080 n^3 - 92667609216 n^2 \\ \hspace{2cm} - 758613219840 n + 918219939840) & n \ge 8
\end{array}
\label{e:16}
\ee
Thus, we are now able to write explicit expressions of all truncated polynomials for arbitrary $n$ up to $m=9$:
\be
\overline{W}_{n,m}(s) = 1 + \sum_{k=1}^{m} A_{n,k}\, s^k
\label{e:17}
\ee
We find numerically the smallest positive roots of  $\overline{W}_{n,m}(-s)$ for odd $m=3,5,7,9$ in order to estimate the values of coefficients in \eq{07b}. It turns out that for all $n$ the observed part of the dependence of the roots on $m$ remains nearly linear. Substituting $m_c=\rho_c n^2$, where $\rho_c$ is the critical density of the jamming transition in the 2D hard-square model, we obtain the estimate of the smallest root $s_*(n)$ of the full polynomial $W_n(-s)$.

\begin{figure}[ht]
\includegraphics[width=16cm]{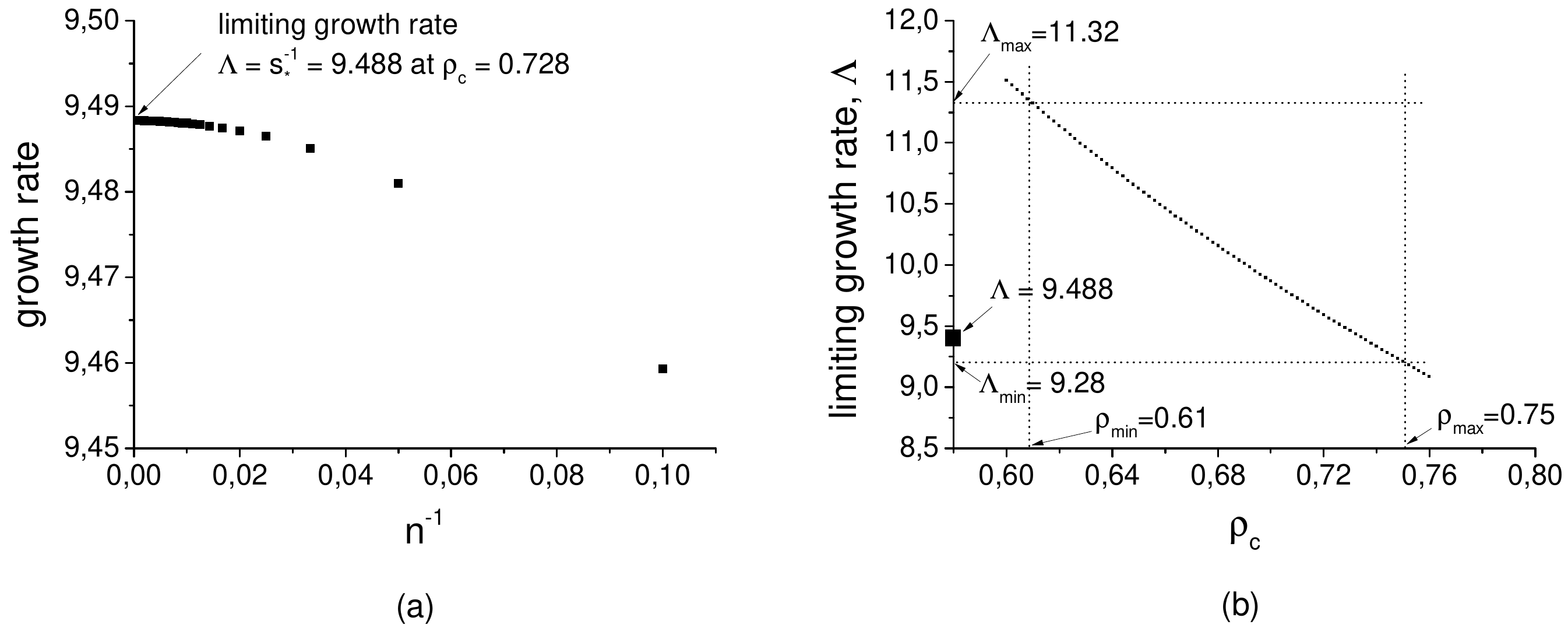}
\caption{(a) Convergence of the growth rate, $\Lambda(n) = s_*^{-1}(n)$, at $n\to\infty$ to the limiting value $\Lambda \approx 9.5$ evaluated at $\rho_c=0.728$; (b) Estimation of limits within which the growth rate $\Lambda$ can vary if $\rho_c$ lies in the interval $0.61<\rho_{c}<0.75$.}
\label{fig:05}
\end{figure}

In \fig{fig:05}a we have demonstrated the convergence of the growth rate, $\Lambda(n) = s_*^{-1}(n)$, at $n\to\infty$ to the limiting value $\Lambda \approx 9.4$ which is evaluated at the transition density in the 2D hard square model, $\rho_c=0.728$. In \fig{fig:05}b we have estimated the limits within which the growth rate of the 3D heap can vary if the transition density $\rho_c$ lies in the interval $0.61<\rho_{c}<0.75$. The limiting growth rate, $\Lambda$, in that case can vary within the interval $[\Lambda_{\min}, \Lambda_{\max}]$, where $\Lambda_{\min} = 9.28$ is evaluated at $\rho_c =0.75$ and $\Lambda_{\max} = 11.32$ is evaluated at $\rho_c =0.61$.

\section{Discussion}

In the paper we provide an estimate of the limiting growth rate, $\Lambda = \lim_{n\to\infty} \Lambda(n)$, of a partition function, $Z_N\sim N^{\theta}[\Lambda(n)]^N$, of 3D heaps of pieces in the large bounding box of base $n\times n$. Our estimate is based on G.X. Viennot's theorem \cite{viennot-rev}, which establishes a relation between the grand canonical partition function of a heap $G_n(s) = \sum Z_N s^N$ and the partition function $W_n(s)$ of distributing pieces in a single layer of a heap. Namely, the theorem states that $G_n(s)=W_n^{-1}(-s)$. The limiting growth rate $\Lambda$ is thereore equal to $\Lambda = \lim_{n\to\infty} s^{-1}_{*,n}$, where $s_{*,n}$ is the closest root of the polynomial $W(-s)$ with smallest absolute value.

In order to estimate the value of roots $s_{*,n}$ we studied the behavior of truncated polynomials $\overline{W}_{n,m}(-s)$, i.e. the polynomials with only first $m$ terms preserved. We noticed that the smallest root of the truncated polynomial grows approximately linearly with growing $m$ up to some  critical value $m_c = \rho_c M_{\max}$, and then saturates. We estimated the linear growth rate by calculating explicitly the coefficients of several first truncated polynomials for arbitrary $n$ and conjecture that $\rho_{c}$ coincides with the ordering transition in the 2-dimensional hard-square gas, which was the subject of a number of classical works and provided estimates of $\rho_{c}$ via plenty of methods, from the perturbation theory and self-consistent approximation, to the mean-field approach and direct numeric computations. The most plausible value of critical concentration is $\rho_{c}\approx 0.728$. That value of $\rho_{c}$ allows us to estimate the limiting growth rate of 3D heaps as $\Lambda\approx 9.5$.

Importantly, the theorem of G.X. Viennot is valid for general heaps of pieces with arbitrary underlying graphs and arbitrary shapes the pieces. Treatment of ensembles of Tetris-like piles of heaps growing in 3D space can be straightforwardly generalized to the consideration of heaps of particles of other shapes, and even to the growth of entangled directed paths, as it is briefly described in \ref{app2}. Statistics of ensembles of non-crossable linear objects with topological constraints has a very broad application area ranging from self-diffusion of entangled directed polymer chains and textures of growing nanotubes, to dynamical and topological aspects of vortex glasses in high-$T_c$ superconductors \cite{nel}.

\begin{acknowledgments}
We are grateful to D. Dhar, M. Bousquet-M\'{e}lou, A. Guttman, G. Oshanin and P. Krapivsky for valuable critique and illuminating comments, to V. Bardakov and A. Malyutin for suggestions concerning 2D extension of locally-free group, and to S. Redner for proposing the transfer matrix approach discussed in Appendix B; S.N. and N.P. acknowledge the support of the Basis Foundation No. 19-1-1-48-1.
\end{acknowledgments}

\begin{appendix}

\section{Generating function of heaps of pieces in a $2\times 2$ and $3\times 3$ bounding
boxes: transfer matrix approach}
\label{app1}

Here, in order to illustrate the statement G. Viennot's theorem, we calculate explicitly the generating functions $G_n(s)$ and $W_n(s)$ for $n=2$ and $3$.

Let us construct a heap layer-by-layer (the ``layer'' is a set of pieces of a heap with fixed vertical coordinate). In each layer the particles-squares must obey the repulsion constraint, i.e. cannot have common sides (although they can have common corners). We start with $n=2$. The maximal possible number of particles in a layer is 2, there are 4 distinct 1-particle configurations (a particle in each of 4 corners) and 2 distinct 2-particle configurations (particles in two opposite corners).
Thus,
\be
W_2(s) = 1 + 4s + 2s^2
\ee
Now, introduce vector $\ve{G}_{2,m}(s)=(G_{2,m}^1(s),G_{2,m}^2(s))^\text{T}$ whose components $G_{2,m}^{1(2)}(s)$ are generating functions of $m$-layer heaps with 1-particle (respectively, 2-particle) configuration in the upper layer. The possible configuration of $m$-th layer depends on the state of $m-1$-th layer. Indeed, it is possible to put 3 different 1-particle configurations  and one of the two 2-particle configurations on top of a 1-particle configurations (those which do not include the opposite corner particle), and it is possible to put any configuration on top of the two particle one. Thus, $\ve{G}_{2,m}(s)$ satisfies
\be
\ve{G}_{2,m}= \ve T_2 (s)  \ve{G}_{2,m-1}
\ee
with transfer matrix
\be
\ve T_{2} = \left(\begin{array}{cc} 3s & 4s \\ s^2 & 2s^2 \end{array} \right)
\ee
and initial condition
\be
\ve{G}_{2,1}= \left(\begin{array}{c} 4s \\ 2s^2 \end{array} \right)
\ee
Thus, generating functions of heaps with arbitrary number of layers and a given configuration on top satisfy
\be
\ve{G}_{2}= \sum_{m=1}^{\infty} \ve T_2^{m-1} (s)  \ve{G}_{2,1} = (\ve I - \ve T_2)^{-1} \ve{G}_{2,1},
\ee
and the desired generating function $G_2(s)$ is
\be
G_2(s) = 1 + G_2^1 + G_2^2,
\ee
where $G_2^1,G_2^2$ are components of $\ve{G}_{2}$. Direct computation gives
\be
(\ve I - \ve T_2)^{-1} = \frac{1}{1-3s-2s^2+2s^3} \left(
\begin{array}{cc} 1-2s^2 & 4s \\ s^2 & 1-3s \end{array} \right)
\ee
and, therefore
\be
\ve{G}_{2}=\frac{1}{1-3s-2s^2+2s^3} \left(\begin{array}{c} 4s \\ 2s^2 - 2s^3 \end{array} \right)
\ee
and
\be
G_2(s) = \frac{1+s}{1-3s-2s^2+2s^3} =\frac{1}{1-4s+2s^2},
\ee
Thus, indeed,
\be
G_2(s) = \frac{1}{W_2(-s)}
\ee
as prescribed by the Viennot's theorem.

\begin{figure}[ht]
\includegraphics[width=13cm]{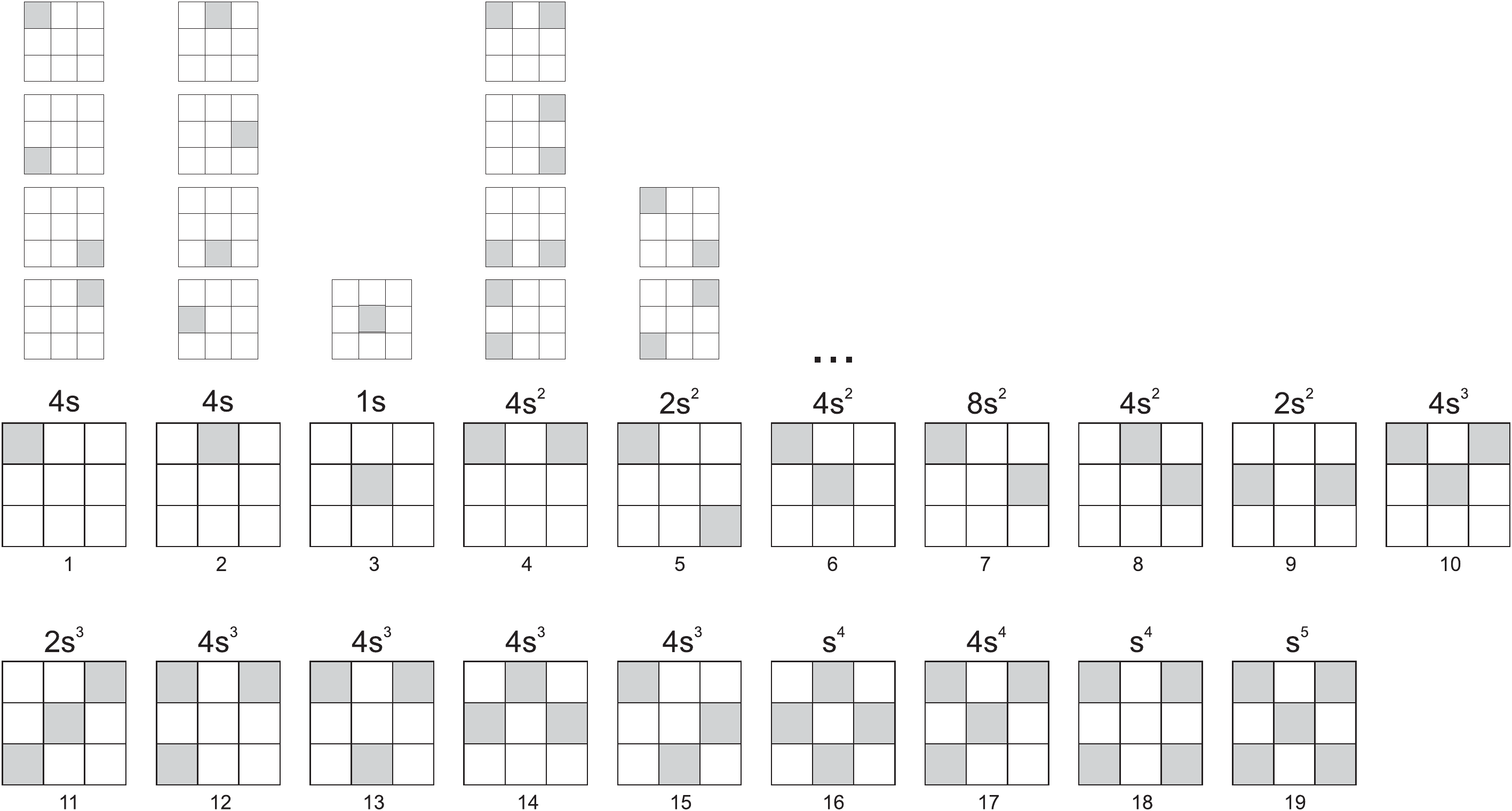}
\caption{Enumeration of different configurations of squares with the hard-core repulsion used in the transfer matrix construction.}
\label{fig:06}
\end{figure}

The corresponding calculation for $n=3$ goes exactly along the same lines but is significantly more cumbersome. Namely, there are 19 topologically different configurations of $1,2,...,5$ ($M_{max}$ is 5 in this case) squares on the $3\times 3$ lattice, as enumerated in \fig{fig:06}. These 19 classes constitute the basis of the transfer matrix. Generally speaking, classes consist of several configurations identical up to rotation or reflection. For example, the topological class 1 in \fig{fig:06} represents 4 configurations of a single particle which could be located in 4 corners of the $3\times 3$ lattice, etc. The corresponding statistical weight of each class is written above the configuration in the figure.

The generating function of a single layer, $W_3(s)$, is the sum of all weights shown in \fig{fig:06}:
\be
W_3(s)=1 + 9 s + 24 s^2 + 22 s^3 + 6 s^4 + s^5
\ee
 (compare \eq{e:07}).

The $ij$-th element of transfer matrix, $T_3$ for heaps in the $3\times 3$ bounding box enumerates the number of ways to put a layer of $i$-th type on top of a layer of $j$-th type. For example, the configuration 1 in \fig{fig:06} can be put on top of:

-- 1 configuration of class 1 (the particle in the previous layer must be exactly below the particle in the current layer);

-- 2 configurations of class 2 (``supporting'' cube in the previous layer can be in one of the two neighboring squares,  but cannot be in one of the two squares along the opposite sides of the box);

-- 0 configurations of class 3 (cental particle in the previous layer cannot support corner square in the current one);

-- 2 configurations of class 4;

-- etc.

The corresponding coefficients $a_{1,1}=1$, $a_{1,2}=2$, $a_{1,3}=0$, $a_{1,4}=2, \dots $ constitute the first row of the matrix $T_3$. Since class 1 consists of a single particle, the coefficients are multiplied by $s^1$. Proceeding row-by-row we obtain a $19\times 19$ transfer matrix $T_3$:
\be
T_3 = \left(
\begin{array}{ccccccccccccccccccc}
 s & 2 s & 0 & 2 s & 2 s & s & 3 s & 3 s & 4 s & 2 s & 2 s & 3 s & 4 s & 4 s & 4 s & 4 s & 3 s & 4 s & 4 s \\
 2 s & s & 4 s & 3 s & 4 s & 4 s & 3 s & 2 s & 2 s & 4 s & 4 s & 4 s & 4 s & 3 s & 4 s & 4 s & 4 s & 4 s & 4 s \\
 0 & s & s & 0 & 0 & s & s & s & s & s & s & 0 & s & s & s & s & s & 0 & s \\
 0 & s^2 & 0 & s^2 & 0 & 0 & 2 s^2 & 2 s^2 & 4 s^2 & s^2 & 0 & 2 s^2 & 4 s^2 & 4 s^2 & 4 s^2 & 4 s^2 & 2 s^2 & 4 s^2 & 4 s^2 \\
 0 & 0 & 0 & 0 & s^2 & 0 & s^2 & s^2 & 2 s^2 & 0 & s^2 & s^2 & 2 s^2 & 2 s^2 & 2 s^2 & 2 s^2 & s^2 & 2 s^2 & 2 s^2 \\
 0 & 2 s^2 & 0 & 0 & 0 & s^2 & 3 s^2 & 3 s^2 & 4 s^2 & 2 s^2 & 2 s^2 & 0 & 4 s^2 & 4 s^2 & 4 s^2 & 4 s^2 & 3 s^2 & 0 & 4 s^2 \\
 0 & 0 & 0 & 2 s^2 & 4 s^2 & 2 s^2 & 4 s^2 & 2 s^2 & 4 s^2 & 4 s^2 & 4 s^2 & 6 s^2 & 8 s^2 & 6 s^2 & 8 s^2 & 8 s^2 & 6 s^2 & 8 s^2 & 8 s^2 \\
 s^2 & 0 & 4 s^2 & 2 s^2 & 4 s^2 & 4 s^2 & 2 s^2 & s^2 & 0 & 4 s^2 & 4 s^2 & 4 s^2 & 4 s^2 & 2 s^2 & 4 s^2 & 4 s^2 & 4 s^2 & 4 s^2 & 4 s^2 \\
 0 & 0 & 2 s^2 & s^2 & 2 s^2 & 2 s^2 & s^2 & 0 & s^2 & 2 s^2 & 2 s^2 & 2 s^2 & 2 s^2 & s^2 & 2 s^2 & 2 s^2 & 2 s^2 & 2 s^2 & 2 s^2 \\
 0 & s^3 & 0 & 0 & 0 & 0 & 2 s^3 & 2 s^3 & 4 s^3 & s^3 & 0 & 0 & 4 s^3 & 4 s^3 & 4 s^3 & 4 s^3 & 2 s^3 & 0 & 4 s^3 \\
 0 & 0 & 0 & 0 & 0 & 0 & s^3 & s^3 & 2 s^3 & 0 & s^3 & 0 & 2 s^3 & 2 s^3 & 2 s^3 & 2 s^3 & s^3 & 0 & 2 s^3 \\
 0 & 0 & 0 & 0 & 0 & 0 & s^3 & s^3 & 4 s^3 & 0 & 0 & s^3 & 4 s^3 & 4 s^3 & 4 s^3 & 4 s^3 & s^3 & 4 s^3 & 4 s^3 \\
 0 & 0 & 0 & 0 & 0 & 0 & s^3 & 0 & 2 s^3 & s^3 & 0 & 2 s^3 & 4 s^3 & 3 s^3 & 4 s^3 & 4 s^3 & 2 s^3 & 4 s^3 & 4 s^3 \\
 0 & 0 & 4 s^3 & s^3 & 4 s^3 & 4 s^3 & s^3 & 0 & 0 & 4 s^3 & 4 s^3 & 4 s^3 & 4 s^3 & s^3 & 4 s^3 & 4 s^3 & 4 s^3 & 4 s^3 & 4 s^3 \\
 0 & 0 & 0 & 0 & 2 s^3 & s^3 & s^3 & 0 & 0 & 2 s^3 & 2 s^3 & 3 s^3 & 4 s^3 & 2 s^3 & 4 s^3 & 4 s^3 & 3 s^3 & 4 s^3 & 4 s^3 \\
 0 & 0 & s^4 & 0 & s^4 & s^4 & 0 & 0 & 0 & s^4 & s^4 & s^4 & s^4 & 0 & s^4 & s^4 & s^4 & s^4 & s^4 \\
 0 & 0 & 0 & 0 & 0 & 0 & s^4 & s^4 & 4 s^4 & 0 & 0 & 0 & 4 s^4 & 4 s^4 & 4 s^4 & 4 s^4 & s^4 & 0 & 4 s^4 \\
 0 & 0 & 0 & 0 & 0 & 0 & 0 & 0 & s^4 & 0 & 0 & 0 & s^4 & s^4 & s^4 & s^4 & 0 & s^4 & s^4 \\
 0 & 0 & 0 & 0 & 0 & 0 & 0 & 0 & s^5 & 0 & 0 & 0 & s^5 & s^5 & s^5 & s^5 & 0 & 0 & s^5 \\
\end{array}
\right)
\ee
Now, similarly to the $n=2$ case the partition function of the heap is
\be
G_3(s) = 1 + \sum_{l=1}^{19} G_3^{(l)}
\label{g3}
\ee
where $G_3^{(l)}$ are elements of the vector
\be
\ve G_3(s) = (\ve I - \ve T_3)^{-1} \ve G_{3,1}(s)
\label{g3vector}
\ee
and $\ve G_{3,1}(s) =(4s,4s,s,4s^2,2s^2,4s^2,8s^2,4s^2,2s^2,4s^3,2s^3,4s^3,4s^3,4s^3,4s^3,s^4,4s^4,s^4,s^5)^{\top}$ is the generating vector of 1-layer heaps. It is easy to evaluate \eq{g3}--\eq{g3vector} using symbol calculation software and find that, indeed,
\be
G_3(s) =\frac{1}{1-9 s+24 s^2-22 s^3+6 s^4-s^5} =\frac{1}{W_3(-s)}.
\ee

\section{Statistics of entangled (2+1)D braids}
\label{app2}

\begin{figure}[ht]
\includegraphics[width=12cm]{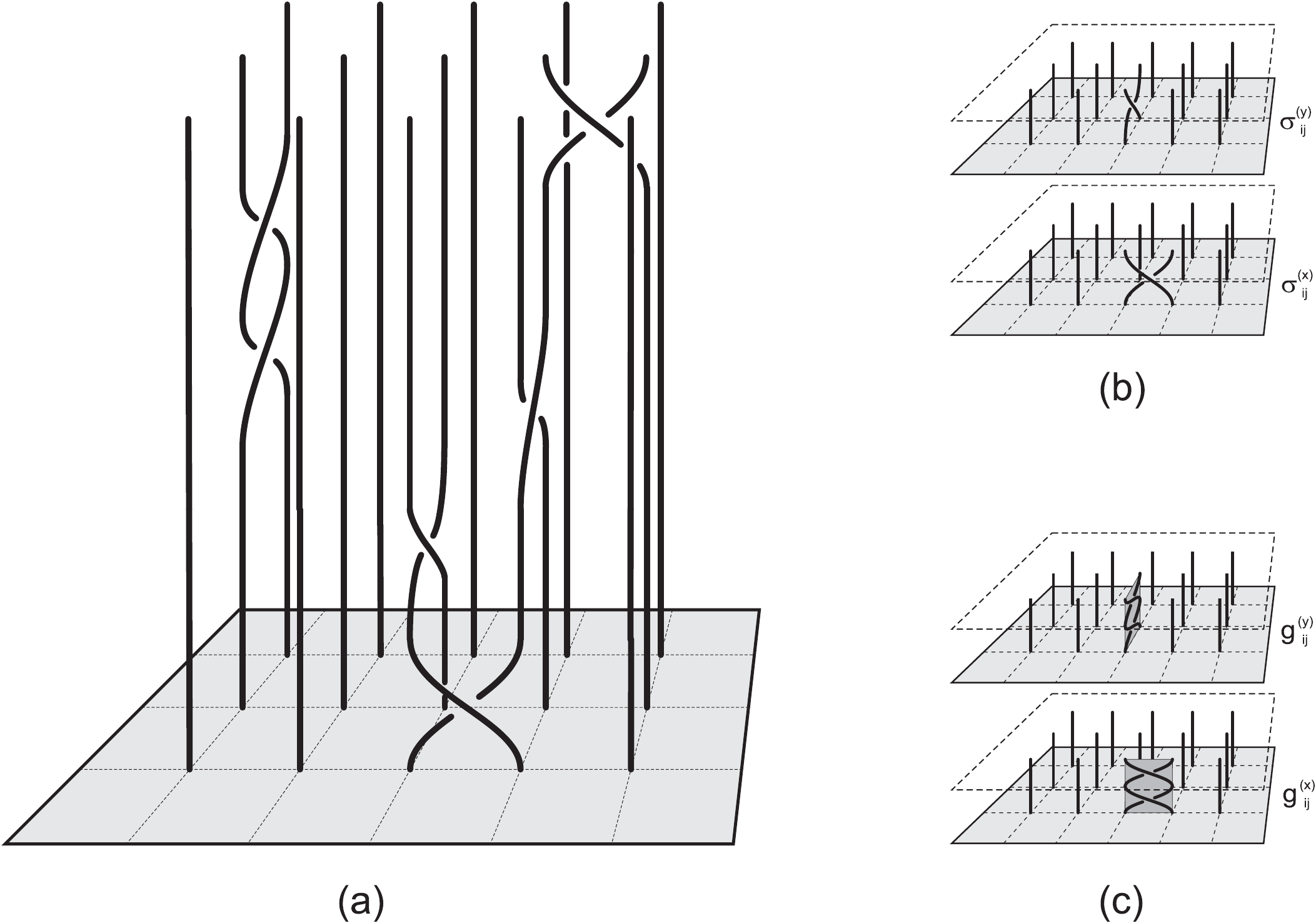}
\caption{(a) Schematic picture of the surface braid in (2+1)D space; (b) surface braid group generator; (c) generator of the ``surface locally free group''.}
\label{2Dbraid01}
\end{figure}

The system shown in \fig{2Dbraid01}a describes the collective dynamics of random directed paths in (2+1)D and ultimately leads to the consideration of a (2+1)D ``surface'' \emph{braid group} $B_n^{2D}$. Following \cite{bir}, consider the two-dimensional lattice $Z^2$ and enumerate its vertices sequentially line-by-line: $P_1, P_2, ..., P_{n^2}$. A (2+1)D-braid of $n^2$ directed strings on $Z^2$ based at $\{P_1, ..., P_{n^2}\}$ is an $n^2$-tuple $b=(b_1,\ldots, b_{n^2})$ of paths, such that:
\begin{enumerate}
\item [(i)] $b_i(1)=P_i$ for $i\in \{1, ..., n^2\}$;
\item[(ii)] $b_i(t)\neq b_j(t)$ for all $\{i,j\}\in \{1,...,n^2\}$, $(i\neq j,\; t\in [1,N])$. The surface braid group $B^{2D}_n$ is the group of homotopy classes of braids based at $\{P_1, ..., P_{n^2}\}$. The group $B^{2D}_n$ has generators $\sigma_{ij}^{(x)}, \sigma_{ij}^{(y)}$, $(i,j)\in \{1,\ldots,n\}$ and their inverses with the standard braiding relations \cite{bir}.
\end{enumerate}
The geometric representation of generators of $B^{2D}_n$ is shown in \fig{2Dbraid01}b.

\begin{figure}[ht]
\includegraphics[width=12cm]{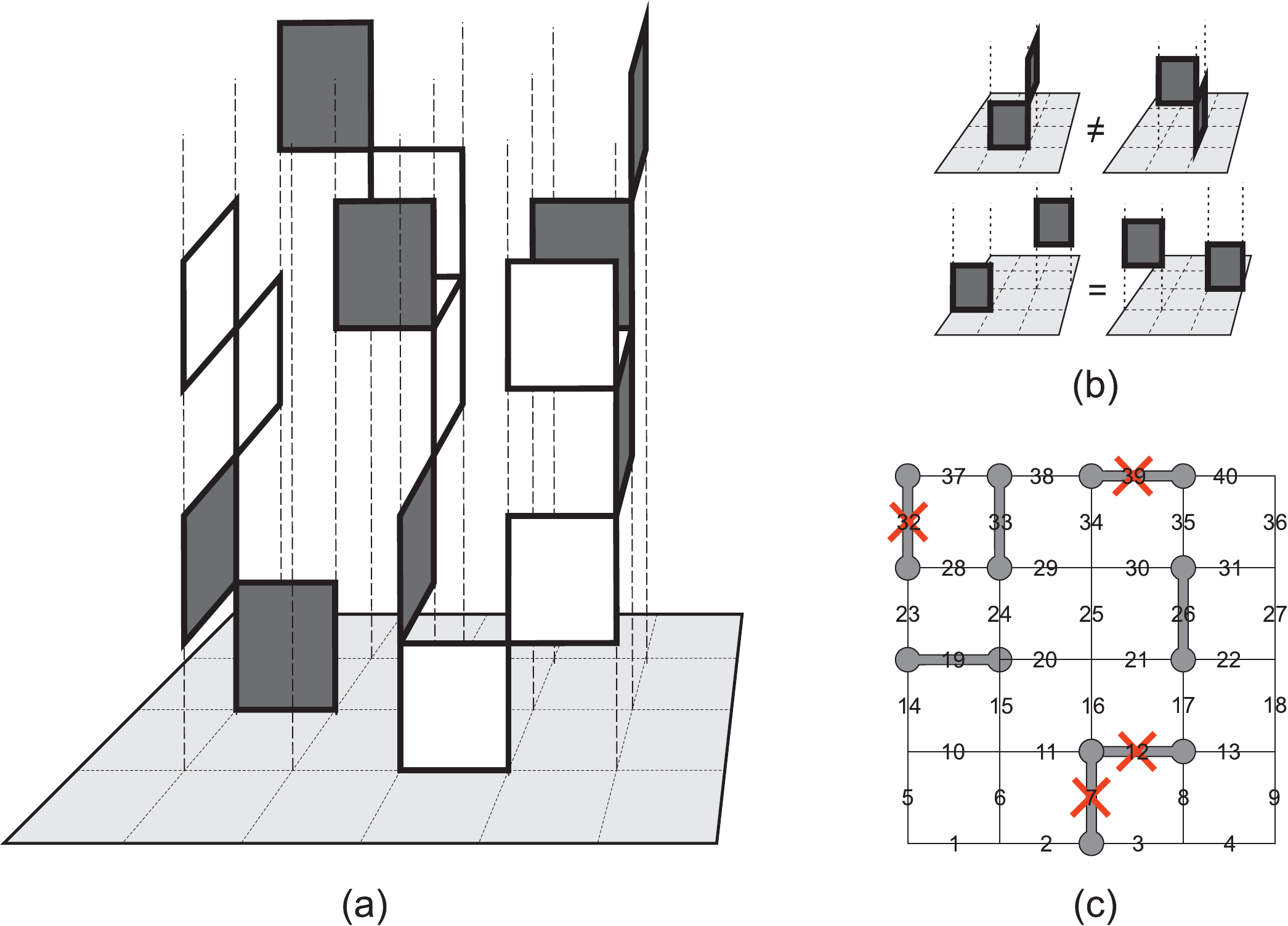}
\caption{(a) Heap consisting of interacting vertically oriented squares. Enumeration of such heaps provides the upper estimate for the number of topologically different configurations of surface braids; (b) generators of surface locally-free group (same as in \fig{2Dbraid01}b); (c) projection of pieces constituting heap to the base plane gibes the collection of dimers with hard core interactions (red crossed demonstrate forbidden configurations).}
\label{2Dbraid02}
\end{figure}

Define now a symmetric random walk on a set of generators (``blocks'') $\{\sigma_{11}^{(x)}, \sigma_{11}^{(y)}, \ldots\}$ with the uniform probability $\frac{1}{n^2}$. This random walk can be viewed as a random uniform ``ballistic deposition'' when we add sequentially new blocks to the roof of a growing heap. One of keynote questions in statistics of entangled lines concerns the evaluation of a number of all nonequivalent topological states after adding $N$ braid group generators. This challenging problem is yet unsolved, however some reasonable estimates can be obtained using the concept of the so-called \emph{surface locally free group} \cite{vershik}. The (2+1)D (``surface'') locally free group ${\cal LF}_{n-1}^{2D}$ is constructed on the basis of the braid group $B_n^{2D}$ by omitting the braiding relations. The group ${\cal LF}^{2D}_{n-1}$ has generators $g_{ij}^{(x)}, g_{ij}^{(y)}$, where $(i,j)\in \{1,\ldots,n\}$, with the following commutation relations (compare \fig{2Dbraid01}c and \fig{2Dbraid02}b):
\be
\left\{\begin{array}{l}
g_{i_1,j_1}^{(x)} g_{i_2,j_2}^{(x)}=g_{i_2,j_2}^{(x)} g_{i_1,j_1}^{(x)}\;
(|j_1-j_2|>0\; {\rm or}\; |i_1-i_2|>1) \medskip \\
g_{i_1,j_1}^{(x)} g_{i_2,j_2}^{(y)}= g_{i_2,j_2}^{(y)} g_{i_1,j_1}^{(x)}\;
(i_2-i_1 \; {\rm or}\; j_1-j_2) \ne \{0,1\}) \medskip \\
g_{i,j}^{(x)}\left(g_{i,j}^{(x)})\right)^{-1}= g_{i,j}^{(y)}\left(g_{i,j}^{(y)})\right)^{-1}=e
\end{array} \right.
\ee

It can be proven that $g_{ij}^{(x)}=\left(\sigma_{ij}^{(x)}\right)^2$ and $g_{ij}^{(y)}= \left(\sigma_{ij}^{(y)}\right)^2$. There is a bijection between words in locally free group and {\it colored heaps}, whose elements are either ``white'' $g_{ij}^{(x,y)}$ or ``black'' $\left(g_{ij}^{(x,y)}\right)^{-1}$. That is, any word written in terms of letters--generators of the group ${\cal LF}^{2D}_n$ represents an unique configuration of a colored heap (see \fig{2Dbraid02}a) in a box of base of $n\times n$ and vice versa: any heap uniquely defines some word in the group ${\cal LF}^{2D}_n$. The configuration of the heap with a black block following immediately after a white one in the same column is forbidden. The process of a heap's growth consists in adding step-by-step new blocks to the roof. The typical configuration of a heap is shown in \fig{2Dbraid02}a.

Thus, the problem of estimating from above the number of different topological states of randomly entangled lines is reformulated as the problem of evaluating the number of different configurations (i.e. the partition function) of a heap raised of $N$ elementary vertical squares representing generators $g_{ij}^{(x)}$ or $g_{ij}^{(x)}$ -- see \fig{2Dbraid02}a and \fig{2Dbraid02}b. To compute the growth of the heap of such pieces, we can use again the relation between generating function of heaps and 2D projection of pieces to the base plane, which are dimers in that case. Thus, we can apply the whole machinery developed for a gas of hard squares to the systems of hard-core interacting dimers on a square lattice (see \fig{2Dbraid02}c) and estimate the growth rate of heap of flat squares in 3D. That will enable us to estimate from above the number of different topological states of randomly entangled directed lines in (2+1)D. We expect to realize the corresponding program in the forthcoming publication.

\end{appendix}


\begin{thebibliography}{99}

\bibitem{cartier} P. Cartier and D. Foata, Probl\`{e}mes combinateires de commutation et r\'{e}arrangements, Lecture Notes in Mathematics (Springer-Verlag: New-York/Berlin), {\bf 85} 1969

\bibitem{viennot-rev} G.X. Viennot, Heaps of pieces, I : Basic definitions and combinatorial lemmas, in: Labelle G., Leroux P. (eds) Combinatoire \'{e}num\'{e}rative, Lecture Notes in Mathematics, (Springer: Berlin, Heidelberg), {\bf 1234} 321 (1993)

\bibitem{hakim1} V. Hakim and J.P. Nadal, Exact results for 2D directed animals on a strip of finite width, J. Phys. A: Math. Gen. {\bf 16} L213 (1983)

\bibitem{viennot1} D. Gouyou-Beauchamps and G.X. Viennot, Equivalence of the Two-Dimensional Directed Animal Problem to a One-Dimensional Path Problem, Adv. Appl. Math. {\bf 9} 334 (1988)

\bibitem{bousquet1} M. Bousquet-M\'{e}lou, New enumerative results on two-dimensional directed animals, Discrete Mathematics {\bf 180} 73 (1998)

\bibitem{bousquet2} M. Bousquet-M\'{e}lou and A. Rechnitzer, Lattice animals and heaps of dimers, Discrete Mathematics {\bf 258} 235 (2002)

\bibitem{vershik} A. M. Vershik, S. Nechaev and R. Bikbov, Statistical Properties of Locally Free Groups with Applications to Braid Groups and Growth of Random Heaps, Comm. Math. Phys. {\bf 212}, 469 (2000)

\bibitem{anim-math} C. Banderier, Ph. Flajolet, Basic analytic combinatorics of directed lattice paths, Theoretical Computer Science, {\bf 281} 37 (2002)

\bibitem{haug} N. Haug, S. Nechaev, M. Tamm, From generalized directed animals to the asymmetric simple exclusion process, J. Stat. Mech. P-10013 (2014)

\bibitem{evans} B. Derrida, M.R. Evans, V. Hakim, and V. Pasquier, Exact solution of a 1D asymmetric exclusion model using a matrix formulation. J. Phys. A: Math. Gen. {\bf 26} 1493 (1993)

\bibitem{betrema} J. B\'{e}tr\'{e}ma and J.G. Penaud, Mod\`{e}les avec particules dures, animaux dirig\'{e}s, et s\'{e}ries en variables partiellement commutatives, ArXiv:math/0106210

\bibitem{bousquet3} M. Bousquet-M\'{e}lou1 and R. Brak, Exactly solved models of polyominoes and polygons, ArXiv:0811.4415

\bibitem{dhar1} D. Dhar, M.K. Phani, and M. Barma, Enumeration of directed site animals on two-dimensional lattices, J. Phys. A {\bf 15} L279 (1982)

\bibitem{dhar2} Equivalence of the Two-Dimensional Directed-Site Animal Problem to Baxter's Hard-Square Lattice-Gas Model, Phys. Rev. Lett. {\bf 49} 959 (1982)

\bibitem{dhar} D. Dhar, Exact Solution of a Directed-Site Animals -- Enumeration Problem in Three Dimensions, Phys. Rev. Lett. {\bf 51} 853 (1983)

\bibitem{baxter} R.J. Baxter, Hard hexagons: exact solution, J. Phys. A: Math. Gen. {\bf 13} L61 (1980)

\bibitem{burley} D.M. Burley, A Lattice Model of a Classical Hard Sphere Gas: II, Proc. Phys. Soc. {\bf 77} 451 (1961)

\bibitem{Fisher65} D. S. Gaunt and M. E. Fisher, Hard‐Sphere Lattice Gases. I. Plane‐Square Lattice, J. Chem. Phys. {\bf 43}, 2840 (1965)

\bibitem{Runnels} L.K. Runnels and L.L. Coombs, Exact Finite Method of Lattice Statistics. I. Square and Triangular Lattice Gases of Hard Molecules, J. Chem. Phys. {\bf 45} 2482 (1966)

\bibitem{Baxter2} R. J. Baxter, I. G. Enting, and S. K. Tsang, Hard-square lattice gas, J. Stat. Phys. {\bf 22}, 465 (1980); R. J. Baxter, Planar lattice gases with nearest-neighbor exclusion, Ann. Combin. {\bf 3}, 191 (1999)

bibitem{pearce} P.A. Pearce and K.A. Seaton, A Classical Theory of Hard Squares, J. Stat. Phys. {\bf 53} 1061 (1988)

\bibitem{blote} G. Kamieniarz and H.W.J. Bl\"{o}te, The non-interacting hard-square lattice gas:
Ising universality, J. Phys. A: Math. Gen. {\bf 26} (1993) 6679

\bibitem{fixman} A. Baram and M. Fixman, Hard square lattice gas, J. Chem. Phys. {\bf 101} 3172 (1994)

\bibitem{mean} H.C. Marques Fernandes, Y. Levin, and J.J. Arenzon, Equation of state for hard square lattice gases, Phys. Rev. E {\bf 75} 052101 (2007)

\bibitem{seq-sol} Wolfram sequence solver, {\verb"https://www.wolframalpha.com/"}

\bibitem{encyclopedia} The On-Line Encyclopedia of Integer Sequences (OEIS), {\verb"https://oeis.org/"}

\bibitem{chess} V\'{a}aclav Kot\v{e}\v{s}ovec, Non-attacking chess pieces (chess in mathematics), 6th edition, {\verb"http://www.kotesovec.cz/books/kotesovec_non_attacking_chess_pieces_2013_6ed.pdf"}

\bibitem{nel} D.R. Nelson, Vortex Entanglement in High-$T_c$ Superconductors, Phys. Rev. Lett. {\bf 60} 1973 (1988); D.R. Nelson, H.Seung, Theory of melted flux liquids, Phys.Rev. B {\bf 39} 9153 (1989)

\bibitem{bir} J. Birman, Mapping class groups of surfaces, Contemp. Math. {\bf 78} 13 (1988)


\end{thebibliography}
\end{document}